\documentclass[preprintnumbers, floatfix,letterpaper,aps,prd,epsfig,nofootinbib,
twocolumn
]{revtex4-1}
\usepackage{bm,graphicx,dcolumn,epstopdf,epsf, latexsym,mathbbol, amssymb,amsmath,color,slashed, mathrsfs,mathcomp,simplewick}
\usepackage[center]{subfigure}
\usepackage{multirow}
\usepackage{makecell}
\usepackage[colorlinks,linkcolor=blue,citecolor=blue,urlcolor=blue]{hyperref}

\begin{document}
\allowdisplaybreaks
 \newcommand{\bq}{\begin{equation}}
 \newcommand{\eq}{\end{equation}}
 \newcommand{\bqn}{\begin{eqnarray}}
 \newcommand{\eqn}{\end{eqnarray}}
 \newcommand{\nb}{\nonumber}
 \newcommand{\lb}{\label}
 \newcommand{\f}{\frac}
 \newcommand{\p}{\partial}
\newcommand{\PRL}{Phys. Rev. Lett.}
\newcommand{\PLB}{Phys. Lett. B}
\newcommand{\PRD}{Phys. Rev. D}
\newcommand{\CQG}{Class. Quantum Grav.}
\newcommand{\JCAP}{J. Cosmol. Astropart. Phys.}
\newcommand{\JHEP}{J. High. Energy. Phys.}
\newcommand{\arXiv}{https://arxiv.org/abs}
\title{Inflationary Cosmology with Quantum Gravitational Effects and the Swampland Conjectures}

\author{Qiang Wu}

\author{Tao Zhu}
\email{zhut05@zjut.edu.cn; Corresponding author}


\affiliation{Institute for Theoretical Physics and Cosmology, Zhejiang University of Technology, Hangzhou, 310032, China}

\date{\today}

\begin{abstract}

Recently proposed two swampland criteria that arising from string theory landscape lead to the important challenge of the realization of single-field inflationary models. Especially one of swampland criteria which implies a large tensor-to-scalar ratio is strongly in tension with recent observational results. In this paper, we explore the possibility if the swampland conjectures could be compatible with single-field inflationary scenarios if the effects due to the quantum theory of gravity are considered. We show that the quantum gravitational effects due to the nonlinear dispersion relation provides significant modifications on the amplitude of both the scalar and tensor perturbation spectra. Such modifications could be either raise or reduce the perturbation spectra depending on the values of the parameters in the nonlinear terms of the dispersion relations. Therefore, these effects can reduce the tensor-to-scalar ratio to a smaller value which helps to relax the tension between the swampland conjecture and observational data.

\end{abstract}



\maketitle

\section{Introduction}

As one of the most promising candidates for a ultraviolet completed description of the quantum gravity that combines gauge and gravitational interactions, string/M theory is expected to provide possibilities for an explicit realization of the cosmological inflationary paradigm. Indeed, at effective level, there are a lot of phenomenological single scalar field inflation models that can arise from the String/M theory. However, in order to consistently embed such single scalar field inflation models into a quantum theory of gravity, it was proposed recently that they have to pass the two criteria of the swampland conjectures \cite{Ooguri:2006in,Vafa:2005ui}. Specifically, the swampland conjectures includes two criteria which states that the scalar inflaton field $\phi$ { being consistent with} a reasonable quantum description of gravity {has} to fulfill the following two conditions
\begin{itemize}
\item The Swampland Criterion I (SCI): The excursion of the scalar field in the field space is bounded by \cite{Ooguri:2016pdq}
\bqn
\frac{|\Delta \phi|}{M_{\rm Pl}} \lesssim c_1 \sim \mathcal{O}(1),
\eqn
\item The Swampland Criterion II (SCII): The gradient of the scalar potential $V(\phi)$ with $V(\phi)> 0$ is limited by \cite{Obied:2018sgi}
\bqn
M_{\rm Pl} \frac{|V_{\phi}|}{V} \gtrsim c_2 \sim \mathcal{O}(1),
\eqn
\end{itemize}
where $M_{\rm Pl}$ is the reduced Planck mass and $V_{\phi} = dV(\phi)/d\phi$. Here $c_1$ and $c_2$ are two positive constants of order unity.

The first criterion is not surprise since it reflects the condition for the validity of the effective field theory of inflation and can be fulfilled by a lot of single scalar field inflation models. For the second criterion, it obviously violates the slow-roll condition, thus leads to a strong tension with the standard slow-roll inflation of the single scalar field inflationary models \cite{Kehagias:2018uem, Agrawal:2018own}, in which the  slow-roll parameter $\epsilon_V$ is defined as
\bqn
\epsilon_V \equiv \frac{1}{2} M_{\rm Pl}^2 \left(\frac{V_\phi}{V}\right)^2.
\eqn
Thus the SCII requires $\epsilon_V \gtrsim \frac{1}{2} c_2^2$ and results in a large tensor to scalar ratio $r \gtrsim 8 c_2^2$ which is obviously inconsistent with precise observational data \cite{Akrami:2018odb, Ade:2018gkx}. This tension has been extensively studied in a lot of recent works, for instance see refs. \cite{Das:2018hqy, Das:2018rpg, Motaharfar:2018zyb, Ashoorioon:2018sqb, Lin:2018kjm, Achucarro:2018vey, Garg:2018reu, Lehners:2018vgi, Dias:2018ngv, Matsui:2018bsy, Ben-Dayan:2018mhe, Kinney:2018nny, Brahma:2018hrd, Kawasaki:2018daf, Dimopoulos:2018upl, Garg:2018zdg, Schimmrigk:2018gch, Yi:2018dhl, Chiang:2018lqx, Holman:2018inr, Bastero-Gil:2018yen, Scalisi:2018eaz} and references therein. More recently, a refined swampland conjecture is proposed, which is \cite{Ooguri:2018wrx}
\begin{itemize}
\item The refined Swampland Criterion (rSCII): the derivatives of the scalar potential $V(\phi)$ are limited by
\bqn
M_{\rm Pl}^2 \frac{|V_\phi|}{V} \gtrsim c_2 \;\;\; {\rm or}\;\;\; M_{\rm Pl}^2 \frac{V_{\phi\phi}}{V} \lesssim - c_3,
\eqn
\end{itemize}
where $V_{\phi\phi}=d^2V/d\phi^2$ and $c_3$ is a third positive constant with order one. This refined version of the swampland criterion is weaker than SCII and its implications on inflation and cosmology have been discussed extensively, see refs. \cite{Wang:2018kly, Fukuda:2018haz, Park:2018fuj, Lin:2018rnx, Chiang:2018lqx, Cheong:2018udx, Kinney:2018kew, Seo:2018abc, Cai:2018ebs} for examples. For rSCI, it is observed that the original second swampland condition SCII now is included in rSCII as only one of possible conditions. It is because of rSCII, some of the single scalar field inflation models could be compatible with the swampland conjectures. However, SCII is still one of possible conditions and in the current research, we will concentrate on it and provide a proposal that could be used to relax its tension with observational data.

In general, CMB temperature anisotropy derived from the inflation models are sensitive to the vacuum state of the perturbation modes. Since the energy scale of inflation at the earlier stage of the inflation is not far from the Planck energy \cite{Martin:2000xs, Brandenberger:2012aj}, one naturally expects that the effects of the quantum gravity can leave some effects on the perturbation modes, which could produce excited initial conditions for the inflationary perturbations. For instance, in loop quantum cosmology, an excited states on the primordial perturbation modes can be generated during the quantum bounce phase prior to the inflation \cite{Wu:2018sbr, Jin:2018wdx, Zhu:2017onp, Zhu:2017jew, Zhu:2016dkn}. {A similar dynamics for quantum bounce can also be achieved in the framework of the effective field theory description of nonsingular bounce \cite{Cai:2014zga}. It is worth noting that the nonsingular bounces from the phenomenological considerations of the effective field theory analysis provides an alternative way to address the initial state issues of the primordial perturbations, see refs. \cite{Cai:2016hea, Cai:2014jla, Cai:2014zga, Cai:2012va} for examples.} In Ho\v{r}ava-Lifshitz theory of quantum gravity, such excited states can be produced by the contribution of high-order spatial derivative terms in the action of the theory, which also supply a nonlinear dispersion relation for the inflationary perturbations \cite{Zhu:2013fja, Zhu:2012zk, Zhu:2011yu}. We note that such nonlinear dispersion relation can also arise from high-order extension of the effective field theory of inflation \cite{Ashoorioon:2018uey,Ashoorioon:2017toq, Ashoorioon:2018ocr, Ashoorioon:2011eg, Ashoorioon:2010xg} and {phenomenological consideration of achieving a nearly scale-invariant power spectrum \cite{cai_primordial_2009}}, for examples.

For the nonlinear dispersion relations, normally it arises from the theory that violates the Lorentz symmetry at the high energy regime. For example, in Ho\v{r}ava-Lifshitz theory of quantum gravity, the Lorentz symmetry has to be violated when the high-order spatial derivative terms dominated at high energy regime, and restores in the low energy limit \cite{Horava:2009uw, Wang:2017brl}. Since the swampland conjectures are based on the analysis that only restricts to the effective theory with Lorentz symmetry, it is important to see if the effects of the Lorentz violation can make the single scalar field inflation models compatible with the swampland conjectures.  In fact, it is proposed very recently that the strong tension between the swampland conjecture SCII and the single field inflationary modes can be relaxed by the excited initial conditions on the perturbation modes \cite{Ashoorioon:2018sqb, Brahma:2018hrd}. As we mentioned, the nonlinear dispersion relation can provide a natural mechanism for generating excited initial states.

In this paper, we consider concrete nonlinear dispersion relations for both the scalar and tensor perturbations and discuss their implications on the swampland conjectures. The nonlinear dispersion relations considered here can be concretely realized in the Ho\v{r}ava-Lifshitz theory of quantum gravity. We show that the nonlinear dispersion relation can modify both the scalar and tensor perturbation spectra but still keep the scale invariance. By using the analytical expressions of perturbation spectra derived from the uniform asymptotic approximation, it is shown that the modification of spectra by the nonlinear dispersion relation can significantly relax the strong tension between the swampland conjectures and the single field inflation.

\section{Effects of nonlinear dispersion relations}

Inflationary theory of the early universe provide a natural mechanism for the generation of the formation of the large scale structure and anaistropies in the cosmic microwave background (CMB) . However, it is still suffering from the trans-Planckian issue considering its energy scale at the earlier stage of the inflation is close to the Planck scale \cite{Martin:2000xs, Brandenberger:2012aj}. To address the trans-Planckian issue, one approach is to consider the nonlinear dispersion relations for both the inflationary scalar and tensor perturbations \cite{Brandenberger:2012aj, Starobinsky:2001kn, Starobinsky:2002rp}.  It is interesting to mention here that the nonlinear dispersion relation can arise naturally from the Ho\v{r}ava-Lifshitz theory of quantum gravity \cite{Zhu:2013fja, Zhu:2012zk, Zhu:2011yu, Horava:2009uw, Wang:2017brl}. Recently it is also shown that such relations can arise from high-order extension of the effective field theory of inflation \cite{Ashoorioon:2018uey}. In this section, we show that the nonlinear dispersion relation can modify both the inflationary scalar and tensor spectra significantly, which could provide a mechanism to relax the tension between the SCII and Planck data.

 To proceed, let us start with the equations of motion for the scalar and tensor perturbations. With the nonlinear dispersion relation $\omega_k^2(\eta)$, the inflationary mode function $u_k(\eta)$ for perturbations  (scalar or tensor) obey the modified Mukhanov-Sasaki equation
\bqn\lb{eom}
u_k''(\eta)+\left(\omega_k^2(\eta)-\frac{z''}{z}\right)u_k(\eta)=0.
\eqn
Here $\eta$ represents the conformal time,  a $'$ denotes derivatives of $\eta$, and $z(\eta)$ is related to the slow-roll evolution of the background. We parametrize the nonlinear dispersion relation in the form of
\bqn\lb{omega}
\omega^2_k(\eta) &=& k^2 \left[1-b_1 \left(\frac{k}{a M_*}\right)^2+b_2 \left(\frac{k}{a M_*}\right)^4\right],
\eqn
where $M_*$ is the relevant energy scale of trans-Planckian physics, $k$ is the comoving wavenumber of the mode, $b_1$ and $b_2$ are dimensionless constants. In the Ho\v{r}ava-Lifshitz theory of quantum gravity, the coefficients $b_1$ and $b_2$ can be related to the coupling constants of the theory \cite{Zhu:2013fja, Zhu:2012zk, Zhu:2011yu, Wang:2017brl}, in which $\hat b_2$ is related to the sixth order spatial derivative terms and $\hat b_1$ is related to the fourth order. In , we require $\hat{b}_2>0$.

{For scalar or tensor modes the equation of motion} described by (\ref{eom}) can be solved analytically by the uniform asymptotic approximation developed in \cite{Zhu:2013upa, Zhu:2013fha}. We would like to mention that  this mathematical method has been applied to the calculations of the primordial spectra in a lot of inflation modes with quantum gravitational effects \cite{Zhu:2014wfa, Wu:2017joj, Zhu:2013upa, Zhu:2013fha, Zhu:2014wda, Zhu:2014aea, Zhu:2016srz, Qiao:2018dpp}, calculations of  quantum gravitational effects of loop quantum cosmology \cite{Zhu:2015ata, Zhu:2015owa, Zhu:2015xsa, Li:2018vzr}, studying parametric resonance during inflation and reheating \cite{Zhu:2018smk}, and derivation of quantization condition in quantum mechanics \cite{zhu_QM}. In the uniform asymptotic approximation, we use the dimensionless variable $y=-k\eta$. Then the equation of motion Eq.(\ref{eom}) can be rewritten as
 \cite{Olver1974,Olver1975}
\bqn\lb{eom58}
\frac{d^2u_{k}(y)}{dy^2}=\Big[g(y)+q(y)\Big]u_k(y),\\
g(y)+q(y)= \frac{z''}{z}- \omega^2_k(y).
\eqn
{This is a second-order ordinary differential equation. Normally it's solution is sensitive to the poles and turning points of $g(y)$ and $q(x)$. In the uniform asymptotic approximation, the function $g(y)$ and $q(y)$ are determined by the behaviors of the corresponding error control function around the poles or turning points \cite{Olver1974, Olver1975, Zhu:2013upa}. For the second-order ordinary differential equation (\ref{eom58}), we find that $g(y)$ and $q(y)$ contains a second-order pole at the origin, i.e., $y =0$. In order to ensure the corresponding error control function of the uniform asymptotic approximate solutions \cite{Olver1974, Olver1975, Zhu:2013upa}, the functions $g(y)$ and $q(y)$ have to be chosen as \cite{Zhu:2013upa, Olver1974},}
\bqn
\lb{gy}
q(y)&=&-\frac{1}{4y^2}, \;\nb\\
g(y)&=&\frac{\nu^2}{y^2}-1+b_1\epsilon_*^2 y^2 -b_2\epsilon_*^4 y^4,
\eqn
{where $\epsilon_*^2=H^2/M_*^2$, with $H$ the Hubble parameter and $z''/z\equiv [\nu^2(\eta)-1/4]/\eta^2$ and $a \simeq - (\eta H)^{-1}$.}

\begin{figure}
\includegraphics[totalheight=2.4in,width=3.4in,angle=0]{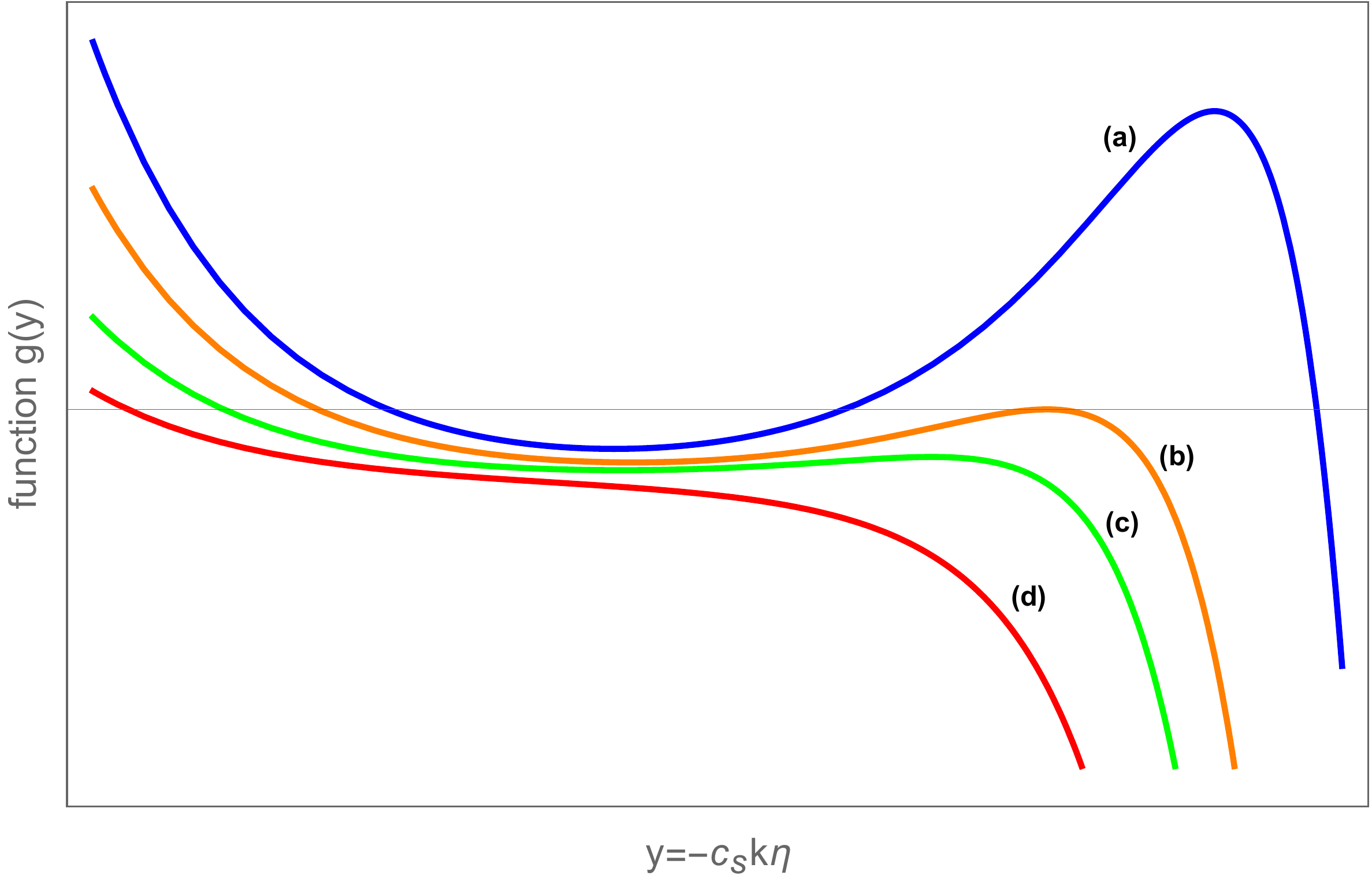}
\caption{The schematic plots of function $g(y)$ in (\ref{gy}) for four representative cases. The number and nature of the turning points for each case are different. Case (a): three single real turning points ($y_0  \ll y_1 < y_2$); Case (b): one single real turning point ($y_0$) and one double real turning point $y_1=y_2 \gg y_0$; Case (c): one single real turning point ($y_0$) and two complex conjugated turning points ($y_1=y_2^*$ and $y_1 \ll {\rm Re}(y_1) = {\rm Re}(y_2) $);  Case (d): one single real turning point $y_0$. Noe that for all cases $b_2>0$. }\label{gofy_tu}
\end{figure}

We observe that the function $g(y)$ defined in the above could also have turning points. {According to} the nature of these turning points, as depicted in Fig.~\ref{gofy_tu}, $g(y)$ can be normally divided into four physical cases \cite{Olver1974}. We label the corresponding turning points of $g(y)=0$ by $y_0$, $y_1$ and $y_2$ with $y_0 < {\rm Re}(y_1) \leq {\rm Re}(y_2)$. For these turning points, $y_0$ is always real, while  $y_1$ and $y_2$ can be both real [c.f. case (a) in Fig.~\ref{gofy_tu}], or both complex [c.f. case (c) in Fig.~\ref{gofy_tu}], or     coalesce into one double root [c.f. case (b) in Fig.~\ref{gofy_tu}].  The case (d) in Fig.~\ref{gofy_tu} corresponds to the case that $g(y)$ has only one single turning point $y_0$. According to the analysis of the uniform asymptotic approximation, \cite{Zhu:2013upa}, the behaviors of the inflationary mode functions are sensitive to the number and nature of the turning points.

With the analysis about the turning points of $g(y)$ in the above, we can employ the uniform asymptotic approximation to construct the corresponding approximate solutions associated about each turning points, which have been presented in details in \cite{Zhu:2013upa}). {By imposing the Bunch-Davies vacuum as the initial state\cite{Zhu:2013upa}, using the approximate solutions of mode function for both scalar and tensor perturbations, the corresponding power spectra can be casted formally in the form} \cite{Zhu:2013upa},
\bqn\lb{pw}
\Delta^2(k)&\equiv &\frac{k^3}{2\pi^2}\left|\frac{u_k(y)}{z}\right|^2\nb\\
&=& \mathcal{A} \frac{k^2y}{4\pi^2z^2\nu}\exp \left(2\int_{y}^{y_0}\sqrt{g(\tilde{y})}d\tilde{y}\right),
\eqn
where $\mathcal{A}$ represents the modification of the power spectra due to the presence of the nonlinear dispersion relation (\ref{omega}), which could be amplified by the non-adiabatic evolution of inflationary perturbations, and is given by
\bqn
\mathcal{A} &\equiv& |\alpha_k+\beta_k|^2\nb\\
&=& (1+2e^{\pi\zeta_{0}^{2}}+2e^{\pi \zeta_0^{2}/2}\sqrt{1+e^{\pi\zeta_0^{2}}}\cos{2\mathfrak{B}}),
\eqn
with
\bqn
\zeta_0^2&\equiv&\frac{2}{\pi}\int_{y_1}^{y_2} \sqrt{g(y)}dy,\\
\mathfrak{B}&\equiv& \int_{y_0}^{\text{Re}(y_1)} \sqrt{-g(y)}dy+\phi( \zeta_0^2/2),
\eqn
{where $\phi(x)\equiv x/2-(x/4)\ln{x^2}+\text{ph}\Gamma(ix+1/2)/2$ with $\text{ph}\Gamma(ix+1/2)$ being the phase of the Gamma function $\Gamma(ix+1/2)$, which is zero when $x=0$, and is determined by continuity otherwise \cite{Zhu:2013upa, Olver1974}. Here $\alpha_k$ and $\beta_k$ denote the Bogoliubov coefficients of the excited state generated by the nonlinear dispersion relation. We see that $\zeta_0$ is related to the integral of $\sqrt{|g(y)|}$ between $y_1$ and $y_2$. When $y_1$ and $y_2$ are two real and single turning points of $g(y)$, $\zeta_0^2$ is positive, while it becomes negative if the the two turning points are complex conjugated. Obviously, the perturbation spectra is amplified by the non-adiabatic evolution of the primordial perturbation since for this case the two turning points are both real and single. When the two turning points are complex conjugated,  since $\zeta_0^2$ is negatively large,  the modified factor $\mathcal{A}$ is order of  $1$ and the violation of the adiabatic evolution of the primordial perturbation is strongly suppressed. 
}

Obviously the perturbation spectra can be modified due to nonlinear dispersion relation which could arise from Ho\v{r}ava-Lifshitz theory of gravity. The effects can be described by two terms. One is determined by the modified factor $\mathcal{A}$ which measures the non-adiabatic effects. Another is due to the exponential integration of $\sqrt{-g(y)}$ from $y_0$ to $0$. To compare different effects, it is convenient to introduce the integral $\mathcal{M}_0$ without the presence of the nonlinear terms in the dispersion relation (by setting $b_1=0=b_2$),
\bqn
\mathcal{M}_0 = \exp{\left(2 \int^{\nu}_y \sqrt{\frac{\nu^2}{y'^2} -1} dy'\right)},
\eqn
When $b_1$ and $b_2$ terms are included, this integral now becomes
\bqn\lb{integral_M}
\mathcal{M} = \exp{\left(2 \int^{y_0}_y \sqrt{g(y')} dy'\right)}.
\eqn
With the help of $\mathcal{M}_0$ and $\mathcal{M}$,  the primordial power spectra (\ref{pw}) can be expressed as
\bqn\lb{PPs}
\Delta_s^2(k) = \mathcal{A}\times \frac{\mathcal{M}}{\mathcal{M}_0} \times \Delta^{2}_{s (\rm GR)}(k),
\eqn
where $\Delta^{2}_{s (\rm GR)}(k) $ denotes the standard nearly scale invariant power-law spectrum when the nonlinear terms in the dispersion relation are set to zero.

To estimate the primordial power spectrum (\ref{PPs}) with the presence of the nonlinear terms in the dispersion relation, let us study the integral in (\ref{integral_M}) in details. For primordial perturbation modes, the inflationary mode function $u_k(\eta)$ for the cosmological scalar perturbation can be related to the comoving curvature fluctuation as $\mathcal{R}^2=\mu^2_k/z^2_s$ with $z_s=\sqrt{2\epsilon} a$, while for tensor perturbation we have $h^2=8\mu^2_k/z_t^2$ with $z_t=a$. In these expression, the Hubble slow-roll parameter $\epsilon$ defined as $\epsilon=-\dot H/H^2$. Then the ratio between the amplitudes of the tensor and scalar perturbation spectra can be calculated via 
\bqn
r&=&\frac{8\Delta_t^2(k)}{\Delta^2_s(k)}=r_{\text{GR}} \left(\frac{\mathcal{A}_t}{\mathcal{A}_s} \sigma_k\right),
\eqn
where $r_{\text{GR}}=16 \epsilon$ denotes the ratio between the amplitudes of the tensor and scalar perturbation spectra predicted in slow-roll inflation models when the nonlinear terms in the dispersion relation are set to zero. The quantity $\sigma_k$ is expressed as
\bqn\lb{sigma-integral}
\sigma_k &\equiv& \mathcal{M}_t /\mathcal{M}_s\nb\\
&=& \exp\left(2\int_y^{y_0^t} \sqrt{g_t(x)}dx-2\int_y^{y_0^s} \sqrt{g_s(x)}dx\right),
\eqn
where the superscript ``s" and ``t" denote the quantities for the scalar and tensor perturbations respectively. We note that we have used $\mathcal{M}_0^{s}/\mathcal{M}_0^t \simeq 1$. In the above expression, we observe that the effects due to the nonlinear terms in the dispersion relation is measured by the factor $ \mathcal{A}_t\sigma_k /\mathcal{A}_s$.

With SCII, we write the ratio between the amplitudes of the tensor and scalar perturbation spectra as
\bqn
r \gtrsim 8 c_2^2 \frac{\mathcal{A}_t}{\mathcal{A}_s} \sigma_k.
\eqn
Considering $c_2 \sim \mathcal{O}(1)$, in order to make the SCII to be consistent with the observational results, one has to impose the following condition,
\bqn\lb{rbound}
\frac{\mathcal{A}_t}{\mathcal{A}_s} \sigma_k \lesssim  {\cal{O}}(0.1).
\eqn
The main purpose of the current paper is to justify that the above criterion can be fulfilled with the presence of the nonlinear terms in the modified dispersion relations.

{From Eq.(\ref{rbound}), for the condition to be satisfied, one can either reduce the modified factor $\mathcal{A}$ or reduce $\sigma_k$. The former possibility is related to the non-adiabatic effects of the primordial perturbations due to the presence of the nonlinear terms in the modified dispersion relations. It is worth mentioning that when we consider the non-adiabatic effects, one assumes $\sigma_k \simeq 1$ for simplicity which can be easily achieved if $\epsilon_*^2 \ll 1$. }

However, once the non-adiabatic evolutions of the primordial perturbations are involved, as we mentioned, the corresponding perturbation modes are non longer at the Bunch-Davies vacuum states and can grows exponentially during the process.  In this case, one has to be at caution about the question that whether the amplification of the non-adiabatic modes could be large enough to destroy the background evolution due to their back-reactions. This important issue has been discussed in details in \cite{Lemoine:2001ar, Brandenberger:2004kx}, which shows that to avoid large back-reactions, the Bogoliubov coefficient $\beta_k$ has to be constrained by
\bqn
|\beta_k|^2\lesssim 8\pi \frac{H^2_{\rm inf}M^2_{\rm Pl}}{M^4_\ast},
\eqn
where $H_{\rm inf}$ is the energy scale of the inflation which is constrained by $H_{\rm inf}/M_{\rm Pl} \leq 2.7 \times 10^{-5}$ due to the most recent Planck 2018 results \cite{Akrami:2018odb, Ade:2018gkx}. Thus, if we take $H_{\rm inf}/M_{\rm Pl} \sim 2 \times 10^{-3} $, one can infer that
\bqn
|\beta_k|^2\lesssim\mathcal{O}(0.1).
\eqn
Then one has
\bqn
\sqrt{1+|\beta_k|^2}-|\beta_k|\lesssim |\alpha_k+\beta_k| \lesssim|\beta_k|+\sqrt{1+|\beta_k|^2},\nb\\
\eqn
which leads to the constraint on $|\alpha_k+\beta_k|^2 $ as,
\bqn
3-2\sqrt{2}\lesssim |\alpha_k+\beta_k|^2 \lesssim3+2\sqrt{2}.
\eqn
Using this constraint, it is obvious that the ratio between the modified factors $\mathcal{A}_t$ for the scalar perturbation and $\mathcal{A}_s$ for the tensor perturbation is restricted to be 
\bqn\lb{bound}
0.03 \simeq (3-2 \sqrt{2})^2<\frac{\mathcal{A}_t}{\mathcal{A}_s}<(3+2\sqrt{2})^2 \simeq 34.
\eqn
This condition provides a strong constraint on the non-adiabatic effects on the primordial perturbation spectrum.  Clearly, from this condition, it is obvious that we have a large space for adjusting parameters $b_1$ and $b_2$ such that ${\mathcal{A}_t}/{\mathcal{A}_s} \lesssim  {\cal{O}}(0.1)$.

\begin{figure}
\includegraphics[totalheight=2.4in,width=3.4in,angle=0]{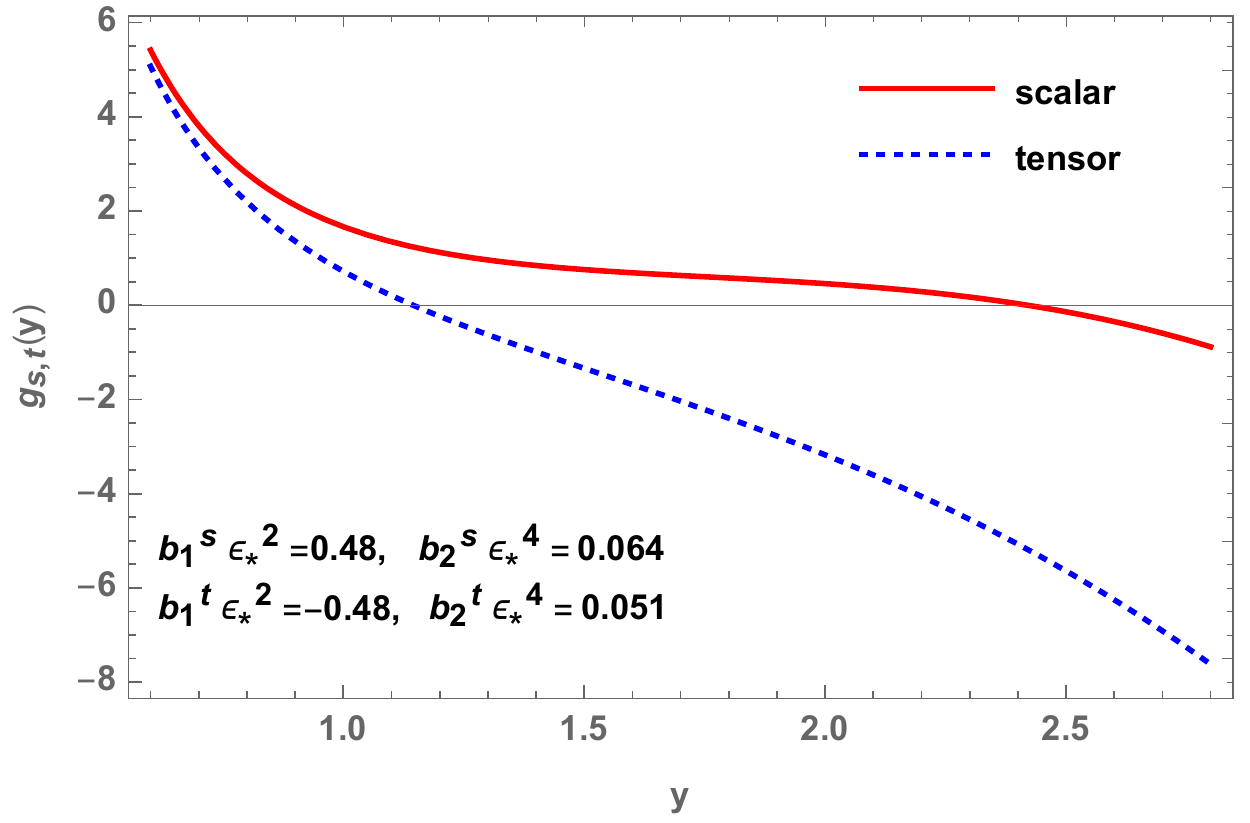}
\caption{ Comparison of $g(y)$ for primordial scalar perturbation and the tensor perturbation in the interval $ y \in (0, y_0)$ for a set  of values for the parameters in the dispersion relation (\ref{omega}).}\label{M}
\end{figure}

{Another way to fulfill the condition (\ref{rbound}) is to reduce the factor $\sigma_k$, which is related to two direct integrals of $\sqrt{g(y)}$ from the turning point $y_0$ until the end of the slow-roll inflation. Therefore, in order to achieve the condition (\ref{rbound}), one has to properly adjust the parameters in the expression of the integrand. As we mentioned, $\mathcal{M}_{s} \simeq \mathcal{M}_t$ when $\epsilon_* \ll 1$, therefore the only way for this to be possible is to relax $\epsilon_* \ll 1$ by requiring $\epsilon_* = H/M_* \lesssim 1$. In order to show the effect of $\sigma_k$ explicitly, we consider $\mathcal{A}_s \simeq 1 \simeq \mathcal{A}_t$. It is worth noting that this implies that the adiabatic condition is satisfied during inflation for the scalar and tensor perturbation modes.} To estimate the integrals in the expression of $\sigma_k$, one observes that due to the nonlinear terms in the modified dispersion relation, the calculation becomes very much mathematics involved. However, for the purpose to show that the condition (\ref{rbound}) can be fulfilled by reducing the value of $\sigma_k$,  we plot the $g_s(y)$ and $g_t(y)$ in Fig.~\ref{M}  by specifying a set of values for the parameters in the dispersion relation. For scalar perturbation we choose $b_1^{(s)}>0$ which leads to a shift of $y_0$ from $\nu$ for linear dispersion relation to a larger value, while for tensor perturbation we consider $b_1^{(t)}<0$ which leads to $y_0 < \nu$. With these reasons, one sees that the curve of $g(y)$ for tensor perturbation is always beneath the scalar one, which implies that $\sigma_k =\mathcal{M}_t / \mathcal{M}_s  <1$. Note that for the purpose to make the SCII to be consistent with observational data, one has to require that $\sigma_k \lesssim \mathcal{O}(0.1)$ and for the parameters chosen in Fig.~\ref{M} we find $\sigma_k \sim 0.1$.

{ Here we would like to make some remarks about the modification on the scalar and tensor power spectra. First, as shown in \cite{Zhu:2013upa, Zhu:2016srz, Qiao:2018dpp}, the effects due to the nonlinear terms in the dispersion relation in the form of (\ref{omega}) can only make modifications on the amplitudes of the primordial scalar and tensor spectrum. This implies that the non-adiabatic evolution of the primordial perturbations due to the nonlinear dispersion relation does not break the nearly scale invariance of the spectrum. Considering the observational data favors a nearly scale invariant scalar spectrum, therefore, the modifications on the power spectrum due to the nonlinear dispersion relation is consistent with the recent observational data. Second, the parameters $b_1$ and $b_2$ involved in the nonlinear terms of the dispersion relation (\ref{omega}) are related to the fourth and sixth order spatial derivative terms in Ho\v{r}ava-Lifshitz theory respectively. While the most of the consistency analysis are related to the parameter $b_2$, the parameter $b_1$ is less constraint. As a result, we have a large parameter space for the parameter $b_1$ that does not lead any inconsistent issues.}

 \section{Conclusions}

 In the current research, we discuss the implications of the quantum gravitational effects due to the nonlinear dispersion relations on relaxing  the strong tension between the recent proposed swampland conjectures and the single field inflationary models. The nonlinear dispersion relations for both the scalar and tensor perturbations considered in this paper can arise naturally in the Ho\v{r}ava-Lifshitz theory of quantum gravity. We show that the quantum gravitational effects due to the nonlinear dispersion relation provides significant modifications on the amplitude of both the scalar and tensor power spectra. Such modifications could be either raise or reduce the power spectra depending on the parameters of the nonlinear dispersion relations. Therefore, these effects can reduce the tensor-to-scalar ratio to a smaller value which helps to relax the tension between the swampland conjecture and Planck data.

\section*{Acknowledgements}
This work is supported by National Natural Science Foundation of China with the Grants No. 11675143 (Q.W. \& T.Z.).



%

\end{document}